# CAHIERS FRANÇOIS VIETE

Série III - N° 11

2021

*Une histoire genrée des savoirs est-elle possible ?*

# De la salle à manger au Collège royal : les espaces savants des collaboratrices en astronomie de Jérôme Lalande

Isabelle Lémonon-Waxin\*

**Résumé**

*Jérôme Lalande, célèbre astronome français au XVIIIe siècle, collabora tout au long de sa carrière, avec plusieurs « calculatrices » en astronomie : Nicole Reine Lepaute, Marie Louise Dupiéry et Marie Jeanne Lefrançois. À la prise en charge de tâches très techniques de calcul et parfois d'observation, elles assumaient également l'intendance scientifique de l'astronome. Cette gestion d'une part de l'entreprise savante s'opérait principalement depuis l'espace domestique, tout comme les calculs astronomiques. Cet espace était donc à la fois espace de vie familiale et espace de production de savoirs. Cet article s'intéresse à son organisation matérielle ainsi qu'aux dynamiques qui s'instauraient entre les différents lieux de savoirs impliqués ici.*

Mots-clés : Lumières, calculatrices, astronomie, espace, femmes, genre, France.

**Abstract**

*Jérôme Lalande, a famous French astronomer in the 18th century, collaborated throughout his career with several female calculators in astronomy: Nicole Reine Lepaute, Marie Louise Dupiéry and Marie Jeanne Lefrançois. Taking on highly technical tasks of calculation and sometimes observation, they also took on the scientific "intendance" for the astronomer. This management of a part of the scholarly enterprise was mainly carried out from home, as were the astronomical calculations. This space was therefore both a family living space and a space for the production of knowledge. This article will focus on its material organization as well as on the dynamics that took place between the different places of knowledge involved here.*

---

\* Centre Alexandre Koyré et Centre de recherche médecine, sciences, santé, santé mentale, société (Cermes3).

# DE LA SALLE A MANGER AU COLLEGE ROYAL …

L'ACTIVITE de productions de tables astronomiques de Joseph Jérôme Lefrançois (1732-1807), également connu comme Jérôme Lalande, est très prolifique au XVIIIe siècle. Elle nécessite le recours à de nombreux calculateurs et calculatrices, lui fournissant un volume de données suffisant à la rédaction de la *Connaissance des temps* ou des *Éphémérides des mouvements célestes*, qu'il dirige pour l'Académie royale des sciences ou la Marine entre 1759 et 1800. Parmi ces calculateurs, trois femmes s'illustrent et sont rendues visibles par les nombreux témoignages de l'astronome : Nicole Reine Lepaute[1] (1723-1788), Marie Louise Dupiéry[2] (1746-1830) et Marie Jeanne Lefrançois[3] (1768-1832). Nous étudierons dans cet article les espaces investis par ces femmes pour pratiquer l'astronomie auprès de Lalande. Le terme espace est ici à entendre dans un premier temps dans son acceptation géographique, et ensuite plus largement dans le sens défini par Michel de Certeau (1980), à savoir un lieu « pratiqué ». Notre intérêt se porte en effet sur l'organisation matérielle de ces « lieux de savoir », régie par les différentes pratiques savantes (dans le temps et l'espace) de ses acteurs et actrices (Jacob, 2014). Notre propos est donc à la fois de circonscrire les interactions entre lieux, savoirs produits, acteurs et pratiques, et d'esquisser la dynamique qui peut s'instaurer entre les différents espaces impliqués dans la production, et la circulation de ces savoirs[4]. En effet, comme l'explique Steven Shapin (1998, p. 6-7), « nous devons comprendre non seulement comment les savoirs sont produits dans des lieux spécifiques, mais aussi comment les échanges s'effectuent entre ces lieux »[5]. Les espaces de production et de circulation participent au façonnage du savoir, tout comme le savoir produit façonne ces espaces[6]. L'appréhension de la production des savoirs s'opère donc conjointement avec celle de leur circulation, et non en opposition. Le cas particulier des pratiques des femmes offre une approche par les marges du fonctionnement des institutions du savoir, et de leurs liens avec des espaces souvent qualifiés trop rapidement de féminins (comme la maison)[7]. En effet, comme l'a montré Volny Fages (2012, p. 23) dans le cadre de son travail sur les communautés de cosmogonistes au XIX$^e$ siècle, cet angle d'étude révèle « l'organisation des institutions scientifiques, et de l'émergence de normes sociales et pratiques en leur sein par la description de ce qu'elles laissent et maintiennent à leur marge ». Nous rejetterons ici les dichotomies qui ont émergé dans les années 1970 dans les pays anglophones lors de la construction de l'histoire des femmes, et qui marquent une frontière très nette entre les associations sphère publique masculine/sphère privée féminine, ou encore espace domestique féminin/espace institutionnel masculin[8]. Cette séparation contribue à naturaliser les espaces selon les catégories de sexe, et il convient donc, à la place, de s'appliquer à identifier les représentations et usages sexués de l'espace pour cerner l'impact des idéologies de genre et de classe sur son organisation. C'est dans cette direction, déjà pointée par Michelle Perrot dans l'*Histoire des femmes* que nous nous inscrivons, en proposant d'examiner les pratiques et usages des espaces matériels des collaboratrices de Lalande, qu'ils soient publics ou privés (Duby & Perrot, 1990-1992). Ces espaces leur sont-ils spécifiques ou partagent-elles ceux des hommes ?

---

[1] Née Étable. Voir (Dumont, 2007 ; Badinter, 2004-2005 ; Boistel, 2004).
[2] Née Pourrat de la Madeleine. Dans cet article, afin de faciliter la recherche plein texte pour le lecteur, l'orthographe Dupiéry attesté à partir de la Révolution est privilégiée, même si avant celle-ci, l'orthographe la plus usitée est Du Piéry (Lémonon-Waxin, 2016 ; Dumont, 2007).
[3] Née Harlay. Elle est également appelée M. J. Lalande, M. J. Lefrançois de Lalande voire Amélie Lalande dans diverses correspondances (Boistel et al., 2010 ; Dumont, 2007).
[4] Cette dynamique est mise en évidence dans (Lémonon-Waxin, 2016).
[5] Ma traduction de : « we need to understand not only how knowledge is made in specific places but also how transactions occur between places ».
[6] Voir (Withers, 2008). L'importance du lieu de savoir dans la production savante est également soulignée dans (Finnegan, 2008 ; Rogers, 2000, p. 189 ; Galison, 1999 ; Ophir & Shapin, 1991).

Pour esquisser une réponse, il est nécessaire de s'intéresser à la structuration, à l'organisation et à la régulation des pratiques savantes dans les espaces matériels qu'elles occupent. L'analyse des modes d'administration des espaces savants pratiqués se fait ici sous l'angle d'une « intendance scientifique »[9] qui dans le cas de l'atelier d'astronomie de Lalande serait dévolue aux femmes. Une intendante scientifique est une femme qui prend en charge, à la place d'un savant, la gestion de certaines affaires en lien avec l'activité scientifique, avec la possibilité de décider par elle-même pour partie d'entre elles[10]. Il s'agit pour elle non de produire des savoirs, mais de participer à leur circulation et/ou d'en faciliter la production, celle-ci étant assurée par un tiers (ici Lalande, oncle et ami)[11]. On peut se demander si les espaces pratiqués par ces femmes sont régis par des normes de genre, et si dans la mise en circulation des savoirs, certains espaces-cibles sont exclus, alors que d'autres sont privilégiés. L'espace domestique[12] sera particulièrement examiné en raison de l'assignation apparente des femmes à cet espace au XVIIIe siècle et sera considéré comme un espace savant[13]. Loin d'opposer espace domestique et espace institutionnel[14], nous montrerons, à l'exemple des travaux proposés dans l'ouvrage *Domesticity in the Making of Modern Science*, comment ces deux espaces collaborent activement à la production des savoirs scientifiques, s'interpénétrant parfois, tout en maintenant une frontière souvent infranchissable pour les femmes, un seuil insurmontable, manifestation sociale du « *threshold* » (« seuil ») domestique au XVII[e] siècle, décrit par Steven Shapin (1988).

---

[7] Rebecca Rogers (2000, p. 189) souligne en effet que « la maison [...] rassemble des espaces masculins et féminins publics », alors même qu'elle est souvent présentée comme un espace féminin privée.

[8] Pour un résumé de l'émergence de l'idéologie des sphères séparées dans les pays anglophones et des oppositions suscitées, voir (Rogers, 2000). En ce qui concerne les notions d'espaces public et privé, on peut rappeler l'influence importante des travaux de Jürgen Habermas (1978), dans le renouvellement du débat entre communication (publicité) et intimité. Nous utiliserons la définition proposée par Antoine Fleury : « L'espace public est un terme polysémique qui désigne un espace à la fois métaphorique et matériel. Comme espace métaphorique, l'espace public est synonyme de sphère publique ou du débat public. Comme espace matériel, les espaces publics correspondent tantôt à des espaces de rencontre et d'interactions sociales, tantôt à des espaces géographiques ouverts au public », Encyclopédie électronique *Hypergeo*, http://www.hypergeo.eu/spip.php?article482, consulté le 15 octobre 2017. Selon cette définition, il apparaît que l'espace privé est constitutif de l'espace public, puisqu'il ne devient privé qu'avec le développement d'une norme sociale de comportement et de sensibilité (Elias, 2016).

---

[9] L'intendance au XVIIIe siècle est la « direction, administration d'affaires importantes » par l'intendant, « celui qui est préposé pour avoir la conduite, la direction de certaines affaires, avec pouvoir d'en ordonner » (Dictionnaire de l'Académie française, 4e et 5e éditions). Son équivalent féminin « intendante » signifie épouse de l'intendant. J'ai choisi d'étendre la définition masculine au cas féminin dans le domaine scientifique (au sens large).

[10] J'ai choisi de ne pas caractériser ces activités sous le terme de la fonction d'assistante ou d'assistant car ce dernier me paraît trop réducteur, présentant les femmes et les hommes concernés dans une démarche de « subissants » plutôt que d'« agissants ». J'ai montré combien leurs activités leur laissent une grande part d'autonomie dans (Lémonon-Waxin, 2019, chapitre 2).

[11] Dans le cas des trois femmes étudiées ici, l'intendance scientifique se superpose à une activité de calculatrices astronomiques.

[12] Par espace domestique, j'entends la maison, le lieu habité. Je ne parlerai pas ici d'espace professionnel, expression anachronique pour le XVIIIe siècle. Le terme « profession » n'apparaît pas avant la fin du XIXe siècle dans le dictionnaire de l'Académie française (7e édition, 1878). Au XVIIIe siècle, l'espace domestique est à la fois espace d'habitation et espace de travail (dans le cadre d'un métier ou non), de production, notamment pour les femmes (Kushner & Hafter, 2015, p. 1-15 ; Tilly & Scott, 1978).

[13] Selon Adi Ophir et Steven Shapin (1991, p. 15), « l'espace savant fixe les conditions d'apparition des objets de la science, de leur validation comme réels, et des conditions dans lesquelles ils sont notables » comme on le constate dans l'espace domestique des savants. Ma traduction de « the place of knowledge lays down conditions for the appearance of the objects of science, for their validation as real, and for the terms on which they are notable ». Pour une analyse approfondie d'un espace savant, se référer à (Lamy, 2007). Voir également la riche approche géographique de David Livingstone (2003). Je remercie vivement Jérôme Lamy pour notre échange sur l'espace savant et pour ses conseils de lecture.

### Des femmes au cœur d'une maisonnée astronomique

Nicole Reine Lepaute, première collaboratrice de Jérôme Lalande en astronomie, selon les archives actuellement disponibles, a semble-t-il, toujours exercé ses pratiques calculatoires dans un espace domestique : le sien bien entendu, mais probablement aussi celui de Jérôme Lalande. En effet, les calculs nécessaires à l'établissement de la table d'oscillations des pendules pour le *Traité d'horlogerie* de son mari (Lepaute, 1755), et le travail d'édition qu'elle a réalisé sur cet ouvrage[15], ont très certainement eu lieu dans l'appartement qu'elle occupait avec Jean-André Lepaute (1720-1789) au Luxembourg, du fait de ses fonctions d'horloger du roi, jusqu'en 1757. Jérôme Lalande qui a reçu les clés de l'observatoire du Luxembourg (dôme nord) en décembre 1754, y loge également, ce qui facilite ses visites chez les Lepaute, où il travaille sur les pendules de l'horloger. Ces pratiques domestiques de production de savoirs, par lesquelles un savant occupant des fonc-tions officielles au sein des institutions savantes, collabore avec des partenaires invisibilisé·es au sein de ces institutions — dont des femmes —, sont très courantes au XVIIIe siècle. Ainsi, Mary Terrall a exposé dans son ouvrage *Catching Nature in the Act: Réaumur and the Practice of Natural History in the Eighteenth Century* (2014) comment à la tête de sa « maisonnée scientifique » (« scientific household »), composée aussi bien d'hommes que de femmes, parents ou non du naturaliste, Réaumur organise sa production scientifique au sein de son espace domestique à l'Hôtel d'Uzès, en s'appuyant sur ces collaborateurs de l'ombre, dont certains sont devenus célèbres et d'autres sont restés totalement inconnus. Ce type de collaboration savante, centrée sur le foyer d'un savant ou de l'un·e de ses collaborateurs ou collaboratrices, est parfaitement illustré par l'atelier d'astronomie de Lalande dont la localisation géographique suit les déménagements successifs de l'astronome et du couple Lepaute.

---

[14] Donald Opitz et ses collègues (Opitz, Bergwick & Van Tiggelen, 2015, p. 2) écrivent vis-à-vis de cette opposition : « Généralement positionnées en termes d'opposition, le professionnel et l'institutionnel vis-à-vis de l'amateur et du domestique, ces cartographies passent à côté des manières plus complexes dont le public et le privé, le professionnel et l'amateur, le civique et le domestique, se sont en fait entremêlés ». Ma traduction de : « *Usually positioned in oppositional terms, professional and institutional vis-à-vis amateur and domestic, such mappings miss the more complex ways in which public and private, professional and amateur, civic and domestic, in fact intermigled* ».

[15] Archives de Genève, Correspondance entre C. Bonnet et J. Lalande, Ms bonnet 26 f 33v, Lettre de J. Lalande du 11 février 1760 : « La muse qui veut bien faire pour moi la connoissance des tems car pour celle qui se fait actuellement je n'y ai que peu de part, est mad. Lepaute, le nom de son mari est celebre par un fort beau traité d'horlogerie, dont on a admiré meme le stile, parce que c'est elle qui avoit présidé à cette partie et vous avés pu voir plus d'une fois ce nom dans les journaux distingués par dessus tous ceux du meme art. » Ici, Lalande ramène Nicole Reine Lepaute à la symbolique masculine de la muse avant même d'évoquer sa production scientifique pour le journal astronomique de la *Connaissance des tems* ou celle de son mari en horlogerie (Lepaute, 1755).

En 1757, la famille Lepaute quitte le palais du Luxembourg pour s'installer dans un immeuble, place de la croix rouge, où Jérôme Lalande demeure aussi à cette époque[16]. Il quitte ensuite ce logement vers 1763, pour la rue Saint Honoré où il réside jusqu'en avril 1771, et investit un appartement place du Palais-Royal, où il a un observatoire[17]. C'est cette même année que le couple Lepaute emménage place du Palais-Royal[18]. Dans les années qui suivent, Lalande confirme sa proximité avec le couple Lepaute dans plusieurs correspondances, citées par Élisabeth Badinter dans son article sur Nicole Reine Lepaute et Lalande (Badinter, 2004-2005, p. 71-76)[19]. Il est clair que les liens entre le couple et l'astronome n'étaient pas que scientifiques, ils étaient également financiers et domestiques[20]. La proximité domestique qui est clairement établie entre 1754 et 1775, année de son installation définitive au Collège royal, est mise en évidence sur la carte présentée en figure 1.

---

[16] Archives nationales, Minutier central des notaires, AN/MC/ET/CI/481, étude Lecourt, Acte du 24 décembre 1757 entre le couple Lepaute et Jérôme Lalande pour un prêt de 3600 livres de l'astronome aux époux Lepaute. Cet acte fait mention de l'adresse des Lepaute et de Lalande à la croix rouge. Je m'appuie ici sur le travail archivistique d'Alain Demouzon que je remercie pour nos échanges, http://www.alain-demouzon.fr/Alain_DEMOUZON_%E2%80%93_site_officiel/Au_fil_du_temps.html

[17] Papiers Boscovich à Berkeley, Lettre de J. Lalande à R. J. Boscovich du 15 avril 1771, cité dans (Badinter, 2004-2005, note 25).

[18] AN/MC/ET/LIII/470, étude Le Pot d'Auteuil, Bail de J. A. Lepaute du 2 octobre 1770.

[19] Les correspondances citées sont les suivantes : Archives de l'observatoire de Paris-Meudon, Ms A-B, 4-10, lettre du 12 janvier 1772 adressée à « M. de la Lalande chez Mrs Lepaute, horlogers du roi, place du Palais Royal » et également une lettre à Jean III Bernoulli du 19 novembre 1773, où Lalande écrit : « le livre de M de Lagrange pour M Caraccioli est encore au logis…. Je l'ai laissé à M Lepaute. ».

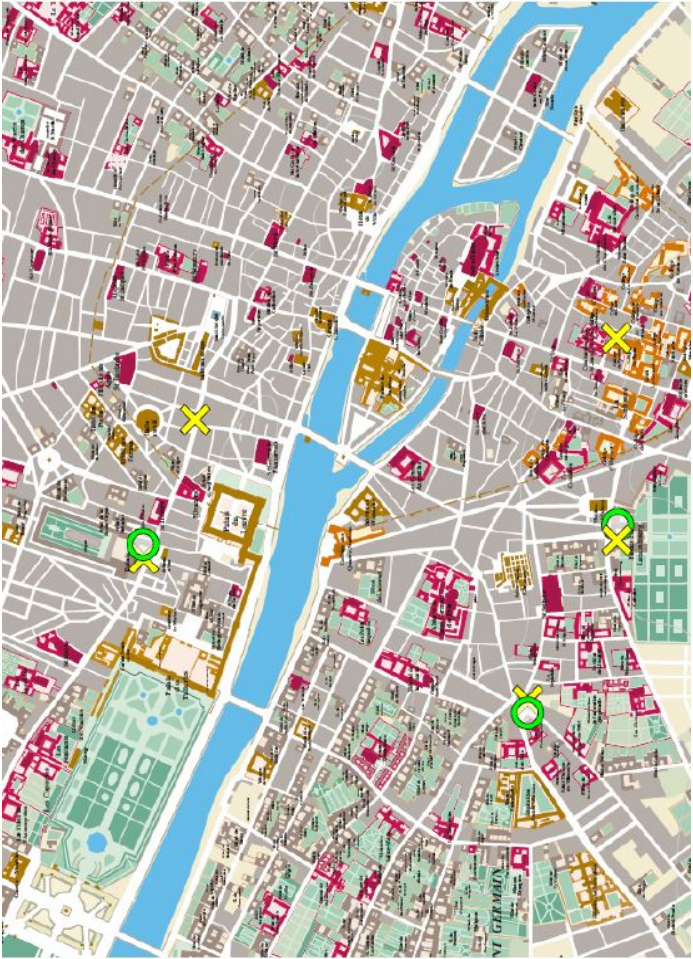

*Figure 1 - Domiciles successifs de J. Lalande (croix jaune) et de N. R. Lepaute (rond vert) à Paris entre 1754 et 1788 ; carte support établie par Michel Huard (Source : http://paris-atlas-historique.fr/, consulté le 13 janvier 2018, Échelle 1/9000e)*

---

[20] Les archives portent la trace de transactions financières entre le couple et l'astronome : voir AN/MC/ET/XXVIII/406, créance au profit de J. A. Lepaute et (Badinter, 2005-2004, note 16).

Il est alors plus facile de comprendre les remarques de Lalande sur le travail qu'il effectue pour Alexis Clairaut avec Nicole Reine Lepaute sur les calculs nécessaires à l'établissement de la trajectoire de la comète de Halley, lors de son retour à proximité de la terre en 1759. L'astronome écrit ainsi, dans sa *Bibliographie astronomique* : « Pendant plus de six mois, nous [Nicole Reine Lepaute et Jérôme Lalande] calculâmes *depuis le matin jusqu'au soir, quelquefois même à table* »[21] (Lalande, 1803, p. 678). Habitant le même immeuble, la collaboration est facilitée, la vie quotidienne de tous s'adaptant au rythme des calculs et inversement, dans un espace domestique organisé certainement pour passer rapidement du repas aux tables de logarithmes, des obligations domestiques aux observations astronomiques ou aux travaux d'horlogerie. Il est très probable que c'est cet espace qui est évoqué de manière sarcastique par Cassini de Thury, rival de Lalande, en 1773 lorsqu'il parle des **«** commis négligents et les ouvriers ignorants d'une manufacture d'astronomie [...] dirigée en second par une académicienne de je ne sais plus quelle académie »[22] (Boistel, 2007, p. 13). Non seulement l'espace domestique des Lepaute-Lalande articule la vie et les activités (domestiques, institutionnelles et de production) du couple, de sa famille[23] et de l'astronome, mais également celles des « calculateurs » de Lalande. En effet, l'astronome emploie de nombreux « techniciens » du calcul astronomique qu'il héberge pour certains en retour (Boistel, 2001, p. 247), et pour lesquels il cherche également une rémunération comme l'indique une copie d'une lettre qu'il rédige le 12 juin 1771 : « il serait bien digne des soins de l'académie royale de marine de faire calculer des tables de la correction de réfraction et de parallaxe dans une forme très simple et très détaillée ; je trouverais ici des personnes pour l'entreprendre si vous aviez pu obtenir quelques fonds pour les aider. »[24] L'immeuble investi en 1771, par le couple Lepaute et Lalande apparaît comme un lieu central d'activité de calculs astronomiques (et certainement horlogers) intimement mêlé à l'espace domestique, comme dans le cas de la maison de Réaumur. Les savoirs produits dans cet atelier domestique circulent d'abord en son sein entre les calculateurs présents physiquement, puis au travers des correspondances de Lalande avec les nombreux astronomes provinciaux et étrangers qui constituent son réseau, pour apparaître par la suite dans les publications institutionnelles ou privées de l'astronome (*Connaissance des temps*, *Éphémérides des mouvements célestes*, *Mémoires de l'Académie des Sciences,* l'*Astronomie*...) ou les journaux savants (*Journal des sçavans*). Ainsi l'espace domestique, opposé parfois à tort à l'espace savant institutionnel, révèle en réalité ici une co-construction des savoirs qui seront par la suite institutionnalisés, par des acteurs et actrices relevant de ces deux espaces. C'est une dynamique de production du savoir instaurée entre le foyer scientifique et l'institution savante, que Monika Mommertz (2005) a également mise en évidence dans le cas de la famille Kirch[25] en Allemagne. Cette circulation entre espaces favorise une co-construction des savoirs où des femmes jouent un rôle important, comme les nombreux cas étudiés dans l'ouvrage d'Opitz et ses collègues l'ont ~~également~~ mis en évidence (Opitz, Bergwick & Van Tiggelen, 2015). En effet, la responsabilité de l'atelier de Lalande est confiée à des femmes en son absence, ainsi en charge de l'intendance scientifique.

---

[21] Jérôme Lalande a écrit ici les calculs réalisés à l'occasion du retour de la comète de Halley. Texte souligné par moi.
[22] Nicole Reine Lepaute est élue associée de l'académie royale de Béziers en 1761. Son nom fait partie de la liste des membres établie en 1766. Archives départementales de l'Hérault, Académie de Béziers, D232.
[23] Jean André Lepaute, arrivé à Paris en 1740, y a été rejoint en 1748 par son frère Jean-Baptiste, et plus tard par certains de leurs neveux pour participer à l'aventure horlogère de la famille.
[24] Columbia University Libraries, David Eugene Smith Historical papers, MN 3605-4, lettre de J. Lalande du 12 juin 1771, destinataire inconnu. Il conseille d'ailleurs cette pratique à ses amis astronomes comme Jacques Potevin (1742-1807) à Montpellier à qui il écrit en 1773 : « si j'étais en province avec le goût que vous avez pour l'astronomie, je prendrais un élève que je ferais observer et calculer : cela coute peu et abat beaucoup d'ouvrage […] j'en ai toujours ici à mes frais, et je m'en suis bien trouvé » (Brossard, 1895, p. 81).

À la suite de Nicole Reine Lepaute, c'est Marie Louise Dupiéry puis Marie Jeanne Lefrançois qui assurent la supervision des travaux des calculateurs[26]. Marie Jeanne Lefrançois, épouse de l'astronome Michel Lefrançois (1766-1839) et donc « nièce »[27] par alliance de Lalande, a vécu dès son mariage en 1788 au Collège royal, lieu de résidence de son époux et de son oncle. Ce lieu institutionnel de la production et de la circulation des savoirs, ne s'est pas réduit à un simple espace de pratiques domestiques pour la jeune femme. En effet, elle se trouve rapidement enrôlée dans l'activité de calculs astronomiques de Lalande et Lefrançois, et multiplie, à l'exemple de Nicole Reine Lepaute, les travaux calculatoires au sein même de son domicile, supervisant les calculateurs de l'astronome lors de ses absences. La double fonction (domestique et productive) du foyer de Lalande, commanditaire des calculs, facilite la participation de ces deux femmes à l'activité scientifique, leur genre préservant l'astronome et ses calculateurs d'une quelconque concurrence institutionnelle. Auraient-elles été acceptées au sein de cette entreprise de production des savoirs, si elles n'avaient pas vécu sur les lieux mêmes de cette production ? Le cas de Marie Louise Dupiéry offre une réponse claire. En effet, lors des voyages de Lalande, et avant le mariage de Marie Jeanne (en septembre 1788) avec son neveu, c'est elle qui est chargée par l'astronome de veiller au bon déroulement de l'activité de ses calculateurs au Collège royal, alors qu'elle vit rue Thévenot (proche de l'actuelle rue Réaumur)[28]. Lalande lui écrit de Londres : « Quand tu auras été au *Collège Royal*, dis-moi ce que la bonne t'aura dit, et M. Barry au sujet de mon Faro ; M. Barry a-t-il bien observé et calculé ? »[29], ou encore : « Je te prie d'aller voir ce qui se passe *chez moi*, de t'informer de ce que fait mon Faro, de lui laver la tête s'il ne travaille pas, de lui demander combien il a calculé d'observations de la Lune » (Lalande, 2007, p. 37 et 41, texte souligné par moi). Ainsi, la participation aux travaux astronomiques au sein du Collège royal d'une femme extérieure à cette institution (y compris d'un point de vue domestique) est attestée.

---

[25] La famille Kirch en Allemagne est l'exemple typique de l'entreprise familiale dédiée toute entière à l'astronomie. Le père Gottfried Kirch (1639-1710), astronome, directeur de l'observatoire de Berlin est secondé pour ses observations et les calculs d'éphémérides, ainsi que le « secrétariat » de son activité d'abord par son épouse Maria Margarethe Winkelmann (1670-1720), puis par ses enfants Christfried (1694-1740) qui reprend la charge de directeur de son père en 1716, Christina (1696-1792) et Margaretha (?-?).

[26] On peut estimer que l'activité de supervision de l'atelier par Nicole Reine Lepaute, qui décède en 1788, après avoir peu à peu perdu la vue, cesse probablement avant 1785. M. L. Dupiéry lui succède jusqu'en 1789, année qui suit le mariage de Michel Lefrançois avec Marie Jeanne, et son installation au Collège royal.

[27] En réalité, Michel Lefrançois n'est pas, comme l'appelle l'astronome, le neveu de Lalande mais son cousin.

L'atelier d'astronomie de Lalande, par sa localisation domestique, facilite l'intégration des femmes à l'équipe des calculateurs de l'astronome. À la fois gestionnaire scientifique de cette équipe (à laquelle elle appartient) et aussi de sa vie domestique, Nicole Reine Lepaute, par sa proximité géographique, offre à Lalande un cadre de travail idéal. C'est une communauté de calcul et de vie qu'elle entretient au service de l'académicien, adaptant probablement durant la journée les espaces dédiés à la production de tables astronomiques aux nécessités de la vie quotidienne (repas, nuit…) dans une forme d'héritage d'une double tradition corporatiste, issue de l'université et des pratiques artisanales[30]. La participation d'une femme, facilitée à pre-mière vue par la localisation du domicile de Lalande ou des Lepaute hors du cadre institutionnel, n'est pourtant pas mise à mal lors de l'installation de l'astronome au Collège royal. Marie Louise Dupiéry y accède semble-t-il, après en avoir obtenu la clé grâce à un passe-droit, pour poursuivre la gestion des calculateurs (Lalande, 2007, p. 46)[31]. Bien sûr, cet investissement d'une savante dans l'atelier situé au Collège, atteint sa plus grande efficacité lorsque Marie Jeanne Lefrançois y intègre le domicile de Lalande après son mariage. Toutes ces techniciennes du calcul astronomique circulent dans les mêmes espaces que leurs *alter ego* masculins, leur présence restant officieuse dans les institutions savantes, et demeurent cependant prisonnières du rôle domestique d'« intendante scientifique ».

---

[28] Marie Louise Dupiéry vit à cette adresse entre les années 1770 et 1793. Elle rencontre Jérôme Lalande en 1779, alors qu'il vit déjà au Collège royal.

[29] Roger Barry (1752-1813) est un lazariste français venu travailler à Paris en astronomie auprès de Jérôme Lalande. Faro est probablement un assistant de Lalande, non identifié pour l'instant. Pourrait-il s'agir d'un surnom donné à son « neveu » Michel Lefrançois, alors âgé de 22 ans ?

[30] Cet héritage dans le champ de l'astronomie est analysé dans (Lémonon-Waxin, 2019, chapitre 2). La transmission et la pratique des savoirs astronomiques aux

### Organiser l'espace domestique pour produire des savoirs astronomiques

Les pratiques savantes dans un environnement domestique entraînent bien souvent une adaptation progressive ou initiale de cet espace. Celle-ci peut se limiter à la simple occupation d'un bureau, cabinet ou chambre, ou s'étendre jusqu'à l'aménagement spécifique de l'espace domestique intérieur voire extérieur. Tout au long de sa « carrière », Lalande se montre particulièrement attentif à la gestion et à l'organisation des espaces savants qu'il occupe, comme le démontrent ses notes sur le Collège royal entre 1776 et 1806[32]. Bien avant sa nomination en tant qu'inspecteur du Collège (1791), il consigne l'état du bâti, les travaux en cours, les répartitions des appartements, les agencements mobiliers et les projets de construction à même d'améliorer le travail savant (dressant un plan à l'occasion)[33]. À titre privé,

---

XVIIe et XVIIIe siècles, sont associées aux traditions artisanales dans (Schiebinger, 1987). On retrouve également cette superposition des activités familiales et scientifiques au sein de l'Observatoire royal de Paris (Deias, 2020, p. 215 ; Ancelin, 2011, p .183-184 ; Kwan, 2010, chapitre 8) ou de l'observatoire de Toulouse (Lamy, 2005, chapitre 2) au XVIIIe siècle.

[31] UR8 lettre du 16 septembre 1791 : « Si l'on a mis mon nom sur ta clé c'est qu'on aura pensé qu'il fallait le nom d'un académicien pour justifier cette exception aux règles... ».

[32] Jérôme de Lalande (1732-1807), « Notes manuscrites de l'astronome Joseph-Jérôme de Lalande sur le Collège royal et ses membres de 1776 à 1806 », *NuBIS*, consulté le 5 janvier 2021, https://nubis.univ-paris1.fr/ark:/15733/1t3t

[33] Dans ses notes, voir en particulier fol. 3 (restauration), fol. 8 (déménagement), fol. 10 (plans pour un projet de bâti), fol. 11 (travaux), fol. 21 (localisation du laboratoire de chimie, travaux dans les logements), fol. 23 (espace pour la bibliothèque), fol. 33 (organisation spatiale de la bibliothèque et du cabinet de

l'astronome consacre une réflexion poussée à l'aménagement de sa maison de Bourg-en-Bresse, lors de la construction de son observatoire (utilisé à plusieurs reprises par Marie Jeanne Lefrançois[34]) qui nécessite la transformation du bâtiment existant en 1792. Lalande établit lui-même plusieurs plans possibles avant la construction, indiquant l'organisation spatiale des pièces (trois séries de ces plans se trouvent encore dans ses archives)[35]. Deux de ces plans comportent une pièce entière réservée à une bibliothèque qui est conservée sur le plan final (voir figure 2 ci-après) où elle occupe une surface de 8 pieds sur 12 (ce qui correspond aujourd'hui environ à 10 m²)[36]. Ce plan précise la localisation de l'observatoire, et un cabinet est présent sur au moins deux des trois plans. Il est clair, que l'organisation spatiale de ce lieu a été pensée soigneusement par Lalande, dans le but d'optimiser au mieux les pratiques savantes partagées avec sa nièce lors de leurs séjours à Bourg. Il a d'ailleurs choisi le point culminant de Bourg-en-Bresse à 242 mètres, pour y faire aménager ce bâtiment, gage d'une vue dégagée sur l'horizon lors des observations astronomiques, auxquelles lui et sa nièce se livrent régulièrement à Bourg. C'est également là que Marie Jeanne Lefrançois effectue en partie les réductions d'étoiles pour le catalogue de l'*Histoire céleste* (Lalande, 1801)[37]. Dans cette demeure familiale, où

---

physique), etc. Le plan présent dans ses notes concerne la configuration d'un terrain situé devant le Collège.

[34] Comme en témoigne cet extrait d'une copie d'une lettre de J. Lalande à Mme Bidal (probable gardienne de la maison) à Bourg-en-Bresse du 12 juin 1798 : « ma fille [Marie Jeanne Lefrançois] ne veut point loüer son observatoire jusqu'à ce qu'elle soit a Bourg c'est a dire au commencement de septembre ». Smithonian libraries, Special Collections (Dibner), MSS 000814 A. Lalande utilise régulièrement dans ses correspondances l'expression « ma fille » pour qualifier Marie-Jeanne Lefrançois, ce qui a amené certains auteurs à écrire qu'elle était sa fille naturelle, ce qui n'est pas le cas (Launay, 2015). Il appelle également Michel Lefrançois, son « fils ». Il renforce le caractère filial de leur relation en attribuant son observatoire à sa nièce, suggérant au passage qu'elle en est la plus grande utilisatrice.

[35] Bibliothèque-Musée Inguimbertine de Carpentras (BMIC), Fonds Raspail, Ms 2762 fol. 170, 171, 176.

[36] Ce plan est associé à une note manuscrite de J. Lalande qui tend à indiquer son caractère final. Il écrit : « le 12 mai 1792 il y a un pied de hauteur. Le 7 nov[embre] 1792 le couvert est fini. » BMIC, Fonds Raspail, Ms 2762 fol. 176.

[37] Dans une lettre écrite de Bourg par J. Lalande, très probablement à J. B. Delambre du 1ᵉʳ octobre (10 vendémiaire), très certainement en 1796 (pendant une grossesse de M. J. Lefrançois) l'astronome écrit : « ma fille est avec moi, elle vous embrasse bien […] elle calcule ici une centaine d'étoiles de 5ᵉ grandeur ou de 5.6 qui n'étaient point dans Flamsteed, vous pouvez juger par la que notre travail

l'astronome et la savante se partagent les pièces au gré de leurs besoins, l'espace est pensé pour favoriser à la fois les pratiques collectives de l'astronomie (au sein de la bibliothèque par exemple, ou dans le jardin pour les observations) et celles qui nécessitent un lieu plus intime (la chambre ou le cabinet) pour les calculs.

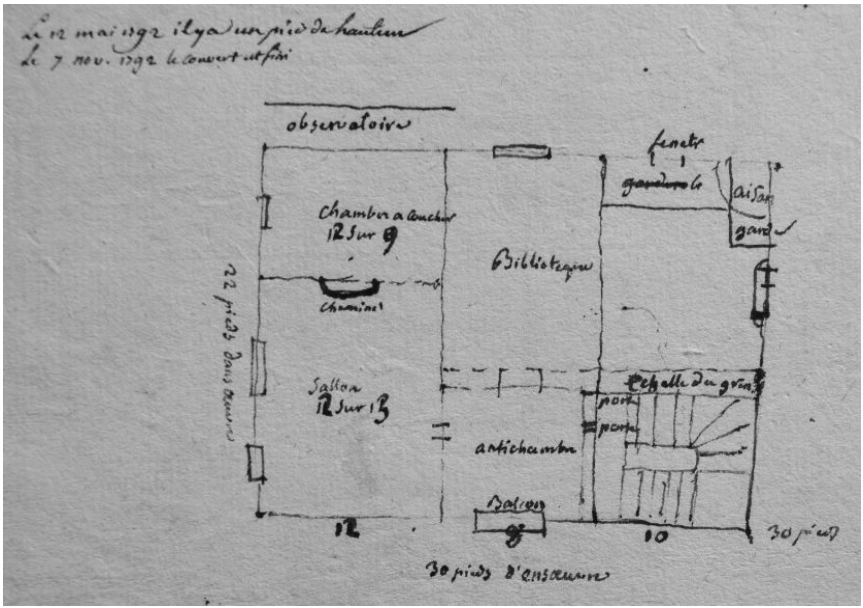

*Figure 2 - Plan final du premier étage de son observatoire à Bourg-en-Bresse de la main de Jérôme Lalande, mai 1792 (BMIC, Archives Raspail, Dossier Lalande, Ms 2762 fol. 170)*

Il en va tout autrement des espaces qu'ils partagent au Collège royal à partir de 1789. En effet, Lalande y partage son logement avec son neveu et sa femme. Élevés entre 1773 et 1783 par Chalgrin, « s'ils sont, dans une certaine mesure des lieux de vie, ces nouveaux bâtiments sont avant tout des lieux d'étude : en leur sein l'espace dédié à la transmission des connaissances occupe une place importante » (Hottin, 2011, p. 24). La priorité est donnée aux espaces d'enseignement (classes, amphithéâtre, bibliothèque, laboratoire ou cabinets), même si le Collège compte neuf appartements, réservés aux plus anciens professeurs, dans les étages. La frontière entre espace institutionnel et espace domestique est clairement marquée comme n'est pas *de minimis.* » Smithsonian Libraries, Special Collections (Dibner), MSS 000814 A. Texte souligné par l'auteur.

on peut le voir sur les plans de Chalgrin (voir figure 3 ci-après). Cependant, l'espace domestique assez réduit des professeurs abrite dans le cas de Lalande un atelier de calculs, au service de l'institution. Marie-Jeanne et Michel Lefrançois, Jérôme Lalande et d'autres calculateurs — comme par exemple l'astronome Burckhardt (1773-1825), un de ses pensionnaires et ami, à partir de 1797[38] — y collaborent dans la pièce de vie principale, constituant également un espace savant.

Marie Louise Dupiéry, quant à elle, n'a pas fait construire un bâtiment où pratiquer les sciences comme l'astronomie ou la chimie, mais a choisi durant la Révolution une maison adaptée à ces besoins savants. Lorsqu'elle quitte Paris pour s'installer à Mareil-en-France en 1793, peut-être pour fuir la Terreur et la vie difficile dans la capitale à cette période, elle achète une maison qu'elle partage avec une autre femme (sa servante ou dame de compagnie)[39]. Elle y poursuit son activité de calculs astronomiques, l'espace disponible y étant propice, et se livre même à des observations dans son jardin où elle installe une méridienne (Lalande, 2007, p. 70). Son inventaire après décès vient compléter la description de son espace domestique, et livre l'aménagement mobilier qu'elle a organisé en lien avec ses pratiques savantes. En effet, Dupiéry dispose parmi tous ses meubles d'une bibliothèque d'environ cinq cents volumes[40], de sphères et de deux mappemondes (une terrestre et une céleste), d'un secrétaire et d'un écritoire, de boites de pastels, de collections d'insectes, de boites remplis de dessins, de coquillages, d'herbiers[41]… auxquels vient s'ajouter un cabinet scientifique (probablement de chimie et/ou minéralogie)[42].

---

[38] Archives de l'Observatoire de Paris, Correspondance de F. X. von Zach à J. Lalande, Ms 1090 fol. 31, Lettre du 23 août 1797.
[39] L'acte d'achat se trouve aux Archives départementales du Val-d'Oise (AD95), 3E29 144V, Étude Gobin (Paris), Vente de maison à Mareil le 6 septembre 1793.
[40] À titre de comparaison, la bibliothèque de l'académicien géomètre J. J. Dortous de Mairan comportait au moment de son décès en 1771, 3400 volumes, et celle de son confrère chimiste A. F. Fourcroy, 2781 titres en 1809. Voir (Roche, 1969, p. 48 ; Fourcroy, 1810).
[41] AD95 2E29 144, Étude Méda (Luzarches), Inventaire après décès établi le 17 mars 1830.
[42] Lettre de J. Lalande à L. B. Guyton de Morveau du 23 janvier 1786. Communication privée (Patrice Bret). Je tiens à remercier Patrice Bret de m'avoir signalé et transmis cette lettre.

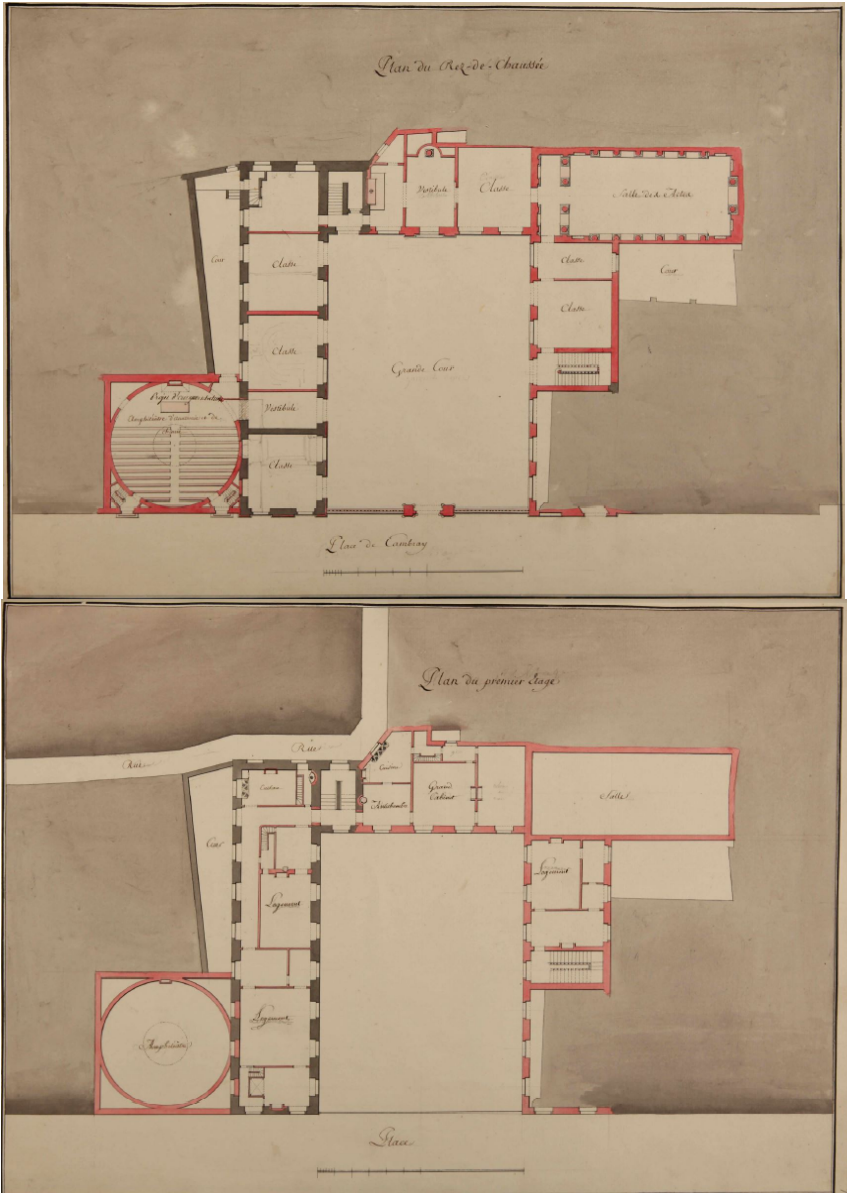

*Figure 3 - Album Chalgrin - Plans, élévations, coupes et profils du Collège Royal de France, dédiés au Roy - 12 Fi - 1774/1780 - 6 - Ark ID : ark:/72507/r20799z7g 0xpsk/f6, plans du rez-de-chaussée et du premier étage © Collège de France. Archives : https://salamandre.college-de-france.fr/archives-en-ligne/ead.html?id=FR075CDF_00Fi0012&c=FR075CDF_00Fi0012_e0000019 - 7 Avril 2021*

Cet ensemble matériel conséquent, lui permettant de réaliser ses calculs astronomiques, comme ses études chimiques, entomologiques ou botaniques, occupe un espace non négligeable de sa maison, dont les plans sont compatibles avec un tel volume d'équipement scientifique (voir figure 4).

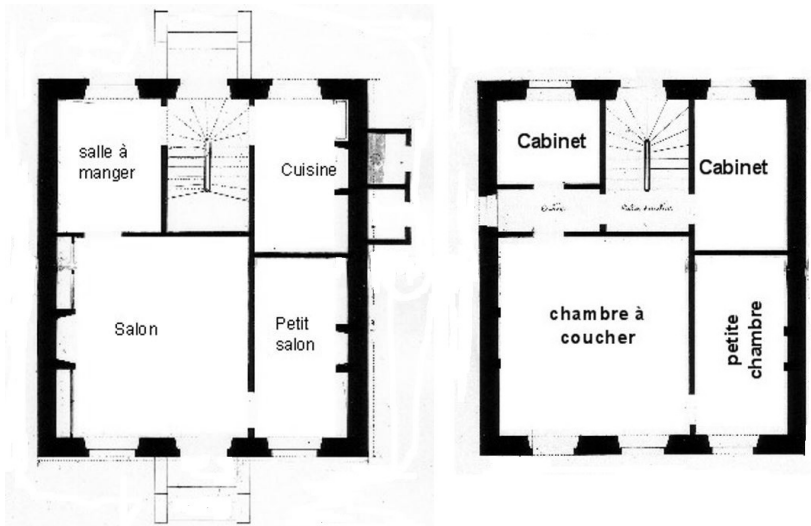

*Figure 4 - Plan du rez-de-chaussée (à gauche) et du premier étage (à droite) de la maison de Marie Louise Dupiéry entre 1793 et 1811 (Archives de la mairie de Mareil-en-France, https://www.mareil-en-france.fr/content/histoire-batiment-mairie)*

Malheureusement, les archives n'apportent pas plus de précision sur l'organisation de son espace domestique, où ont été réalisés certains des calculs nécessaires à la rédaction des tables astronomiques publiées dans la *Connaissance des temps* (1791) ou les *Éphémérides des mouvements célestes* (1783), et les travaux préparatoires à la rédaction de la table analytique de chimie pour Fourcroy (Dupiéry, 1802). Il apparaît que l'espace domestique de Dupiéry a été aménagé pour favoriser son travail savant. Lieux d'intimité (chambres, cabinets) et espaces collaboratifs[43] (bibliothèque, méridienne) s'y côtoient, tout en offrant à la savante le matériel nécessaire à ses pratiques scientifiques (sphères, mappemonde, cabinet de chimie...). À nouveau ici, les productions savantes qu'elle réalise dans son espace domestique sont partagées avec les savants, qui les utilisent dans leurs publications (Lalande, Fourcroy), malgré son éloignement géographique à partir de 1793. Ce dernier n'est plus compatible avec son ancien rôle d'« intendante scientifique » auprès de Lalande, mais ne l'empêche pas de réaliser des tâches données par les savants qu'elle côtoie. La localisation spatiale d'une femme près du domicile d'un savant est donc cruciale pour une employabilité optimum et durable. Celle-ci peut même s'avérer source de nouvelles fonctions « officielles » au sein des institutions savantes, telles que le Collège de France, comme nous le verrons dans le cas de Marie Jeanne Lefrançois.

---

[43] Jérôme Lalande mentionne à plusieurs reprises dans ses correspondances la possibilité d'aller lui rendre visite à Mareil, et il est probable qu'ils aient pu observer et calculer ensemble à Mareil. Voir (Lalande, 2007, p. 53-55), UR15 Lettre du 16 octobre 1793, UR16 Lettre du 19 juin 1794, UR17 Lettre du 3 juillet 1794.

**Le Collège royal, un espace mixte, facteur d'opportunités pour des femmes dans le monde savant**

L'accès du Collège royal à quelques femmes liées à des savants du Collège, par des liens familiaux ou autres, leur permet de s'investir dans le fonctionnement de cette institution. L'espace initialement domestique, que constitue le logement de Jérôme Lalande et du couple Lefrançois, permet à Marie Jeanne, par ses interactions avec les membres du Collège (devenu Collège de France en 1795) et ses allers et venues au sein des bâtiments, d'occuper un rôle de confiance pour l'institution elle-même, au moins dès 1796. Elle est en effet chargée de récupérer l'argent destiné au paiement des professeurs auprès de la trésorerie nationale et de gérer les comptes. Ainsi lors de l'assemblée des professeurs, Lalande note sur le registre le 29 février 1796 : « Lalande a annoncé qu'il a recu le mandat pour le payemen octodecuple des 3 derniers mois 344400#. La Cit[oyenne] Lefrançois ira demain »[44], ou encore en 1797 : « […] La depense a monté à 220# que la Cit[oyenne] lefrancois a payées sur l'argent du bois »[45], et « …on annonce un payement de la trésorerie. La citoyenne Lefrançois s'est chargé de faire les comptes [sic]. »[46] Cet arrangement officieux entre les professeurs du Collège et Marie Jeanne Lefrançois par l'entremise de Jérôme Lalande, inspecteur de cette institution, prend très certainement une forme plus officielle en

---

[44] Archives du Collège de France (ACF), Assemblée des professeurs - Registre des délibérations du 29 février 1796. Le symbole # signifie livres. Le registre de cette assemblée recueille le compte-rendu des réunions et les décisions qui y sont prises.
[45] ACF, Assemblée des professeurs - Registre des délibérations du 15 novembre 1797.
[46] ACF, Assemblée des professeurs - Registre des délibérations du 30 novembre 1797.

1798 avec la mise en place d'une procuration[47] auprès de la trésorerie nationale, et la validation de cette charge par l'assemblée des professeurs. Il est indiqué dans le registre des délibérations en date du 30 mars 1798 : « Lalande pour ne pas retarder les payements a demandé une procuration pour la citoyenne lefrancais qui veut bien prendre la peine d'aller à la Trésorerie. Il propose à la Compagnie de choisir telle autre personne ; qu'elle jugera à propos. L'on demande que la Cit[oyenne] Lefrancais veuille bien continuer. »[48] Cette confiance, associée au crédit dont bénéficie la savante[49] de la part des membres du Collège de France, est renouvelée par les professeurs en 1799, qui notent : « L'affaire de l'emprunt d'Angleterre[50] n'étant pas encore terminée, on prie le C[itoyen] Lalande de s'en occuper incessamment ou par lui-même ou par l'entremise de la C[itoyenn]e françois. »[51] C'est parce qu'elle habite au Collège et qu'elle bénéficie de la caution morale de son oncle Lalande, que Marie Jeanne Lefrançois a l'opportunité d'être employée d'abord officieusement à la gestion financière d'une part de l'institution. La zone d'exercice de son intendance, initialement limitée à l'espace savant de Lalande et Lefrançois, son époux, s'étend. Le crédit grandissant que lui accordent les savants du Collège lui permet par la suite de s'imposer comme une personne de confiance, apte à prendre en charge officiellement (par le biais d'une procuration) une partie des finances de l'institution savante. C'est bien entendu un gain de temps pour les membres du Collège de France qui se déchargent ainsi d'une gestion financière qu'ils auraient dû confier à l'un d'entre eux en l'absence de la savante, mais aussi une forme de reconnaissance de son crédit. La gestion d'une petite partie des comptes du Collège l'intègre à l'administration de l'institution, révélant ainsi sa capacité d'agir (*agency*) dans le monde savant.

---

[47] Cette pratique relativement courante au XVIIIe siècle qui consiste pour un homme à confier la gestion de son patrimoine (quelle qu'en soit la forme) à une femme par procuration a été étudiée entre autres dans (Ferland & Grenier, 2013).

[48] ACF, Assemblée des professeurs - Registre des délibérations du 30 mars 1798. Cette procuration n'a pour l'instant pas été retrouvée. Ce compte-rendu distancié des décisions prises par l'assemblée des professeurs suggère une unanimité de ces derniers quant à la prolongation du rôle confié à Marie Jeanne Lefrançois.

[49] Concernant les notions de confiance et de crédit dans le monde savant du XVIIIe siècle, voir (Haru Crowston, 2013 ; Fontaine, 2008 ; Muldrew, 1998). Recension historiographique dans (Lilti, 2015).

[50] Le 16 nivôse an VI (5 janvier 1798) l'emprunt national pour la descente en Angleterre est ouvert par le Directoire, puis clôturé le 3 nivôse an VII (23 décembre 1798). Les citoyens français sont ainsi appelés à soutenir financièrement l'effort de guerre contre l'Angleterre. Cette clôture entraîne le remboursement des investisseurs, dont faisaient peut-être partie certains membres du Collège de France.

[51] ACF, Assemblée des professeurs - Registre des délibérations du 19 mai 1799.

Isabelle LEMONON-WAXIN

Après avoir examiné la manière dont quelques femmes investissent, transforment ou encore exploitent l'espace pour mettre en œuvre leurs pratiques scientifiques et d'intendance, et intégrer le monde savant, dans un cadre principalement privé, nous interrogerons le rôle de l'espace, et les contraintes qu'il subit, lorsque des femmes s'attèlent à la transmission des savoirs astronomiques, que ce soit en privé ou en public.

**L'espace domestique et la transmission des savoirs astronomiques : de la famille à un public féminin**

Jusqu'au début du XIX$^e$ siècle, l'apprentissage de l'astronomie se fait très souvent sur le mode de transmission des savoirs de l'artisanat et/ou des corporations, d'un maître à son apprenti. C'est par cet apprentissage, en grande partie domestique au domicile du maître astronome, que Jérôme Lalande a étudié l'astronomie aux côtés de Joseph Nicolas Delisle (1688-1768). Quelques archives conservées témoignent également d'un tel investissement de la part des femmes dans leur foyer. Ainsi le couple Lepaute restant sans enfant, il fait venir un de ses « neveux » Joseph Lepaute (1751-1788) qui sera appelé plus tard Dagelet. Lalande (1803, p. 708) raconte que « Mme Lepaute, voyant que j'avais besoin d'un élève astronome, le fit venir à Paris, où il arriva le 25 février 1768, alors qu'il avait été initialement formé en horlogerie par son frère aîné en province ». L'éducation en astronomie donnée par Nicole Reine Lepaute et par Lalande dans l'espace de l'atelier d'astronomie, permet rapidement à Dagelet d'observer et de réaliser des calculs astronomiques avec l'astronome, puis d'en devenir lui-même un de renom. La savante se charge également de l'éducation astronomique d'au moins un des fils de son beau-frère Jean Baptiste Lepaute (1727-1802), horloger, comme le confie Lalande :

> M. Lepaute le jeune, qui fut horloger du roi, puis le chef de la famille, eut des enfants ; l'aîné [Louis Alexandre (1763-1845)] fut élevé par Mme Lepaute avec un soin extrême. […] À l'âge de six ans, il faisait des calculs astronomiques ; et comme il a maintenant trente-sept ans, il serait connu dans les sciences, comme d'Agelet [Dagelet] son cousin, si on lui eût fait embrasser cette carrière[52] (Lalande, 1803, p. 678-679).

---

[52] Dagelet est effectivement le seul membre de la famille Lepaute, hormis Nicole Reine, à s'être investi en astronomie.

Marie Jeanne Lefrançois s'est également illustrée dans la transmission des savoirs astronomiques dans son espace domestique, à l'exemple de Nicole Reine Lepaute. En effet, il est probable qu'elle se soit chargée d'une partie de l'instruction de ses enfants, même si certains ont été placés en pension. En 1804, le savant hollandais Jan Hendrick van Swinden (1746-1823) écrit à la savante :

> Quant à votre fils ainé [Isaac (1789-1855)] je ne doute pas que le gout des Mathématiques ne lui vienne : il est entouré de bons exemples ; ceux-ci exciteront son émulation. […] On me mit peu après [14 ans] au Mathématiques, et celles-ci remplacèrent bientôt la poupée et les chevaux de bois. Il en sera j'espère de même de votre fils : qui d'ailleurs peut recevoir formellement sous mille formes différentes des instructions domestiques, qui dans ce genre me manquent absolument. (van Swinden, 1804, BMIC, Fonds Raspail, Ms 2763 fol. 306-307)

Il est possible que le jeune homme de quinze ans se soit montré rétif aux mathématiques, nécessaires aux calculs astronomiques, malgré un apprentissage domestique. Jérôme Lalande (2014, p. 299) précise cependant deux ans plus tard dans une de ses lettres à l'astronome allemand Johann Elert Bode (1747-1826) : « Mon neveu Isaac commence déjà très bien à calculer et observer pour prendre la place de son père. » Marie Jeanne Lefrançois, son mari Michel et Jérôme Lalande ont probablement tenté d'inculquer à leurs enfants et petits neveux des notions d'astronomie, mais aucun d'entre eux n'a poursuivi dans cette voie. Cette transmission familiale domestique de l'astronome et de son épouse à leurs enfants, quel que soit leur genre, est également identifiée au XVIII$^e$ siècle dans la transmission des savoirs astronomiques nécessaires à la confection des almanachs dans la famille Kirch, en Allemagne (Mommertz, 2005 ; Schiebinger, 1987).

Mais quelques femmes durant les Lumières ne restreignent pas leur enseignement astronomique, et plus largement scientifique, au cadre privé et investissent parfois la sphère publique. Pour autant, ces pratiques, assez rares, s'accompagnent de la mise en scène de leurs corps afin de préserver conventions sociales genrées et réputation. Ainsi, lors de la présentation publique de ses thèses par Laura Bassi (1711-1778) à Bologne en 1732, la cérémonie est adaptée, tant du point de vue spatial (salle différente, accompagnement de l'impétrante par deux femmes de haute noblesse) que social (Bassi n'offre pas des cadeaux comme le veut la tradition pour les hommes, mais en reçoit ; ces présents sont de nature différente) en raison de son sexe et la fait apparaître comme une vierge de sagesse et la Minerve de Bo-logne[53]. Il en est de même pour les cours de mathématiques dispensés par Marie Anne Pigeon (1724-1765) dans la salle de conférence d'André Pierre de Prémontval (1716-1764), son maître, dans les années 1740. Tout est pensé pour préserver sa réputation et gommer toute concurrence menaçante : de son entrée dans la salle, la première, à sa position en bas de l'estrade de manière à tourner le dos au public, jusqu'au rempart des corps des « vieillards » assis auprès d'elle (Lémonon-Waxin, 2019, p. 247). Dans cette pratique publique d'enseignement des sciences, l'agencement spatial de l'oratrice et du public est contraint par la bienséance du XVIII$^e$ siècle, et joue un rôle non négligeable dans l'acceptation de Marie Anne Pigeon d'Osangis en tant que « Maître » de mathématiques par les étudiants. Dans le cas de Marie Louise Dupiéry qui donne des cours publics d'astronomie en 1789 et 1790, la question des conventions et de la réputation semble moins aiguë puisqu'à la fois elle est veuve (et non célibataire à l'instar des femmes précédemment citées), et que ces cours s'adressent « principalement aux dames »[54], ce qui ne peut donner lieu à « calomnie ». Elle utilise pour cela son espace domestique, un appartement parisien, situé au 8 rue Thévenot, ce qui limite la taille du public. Ce cours présente une partie « théorique » qui peut se donner dans un logement suffisamment grand pour posséder un salon, ce qui est le cas chez cette savante. En effet, l'inventaire après décès de son mari réalisé en 1780 à cette même adresse décrit avec précision son espace domestique. Il est composé d'une cave, d'une cuisine, d'un cabinet, de deux chambres et d'une grande salle[55]. Cette salle possède une cheminée surmontée de deux grands miroirs, une tenture, un total de seize fauteuils cabriolets et de deux bergères ainsi qu'un secrétaire en bois de palissandre. Elle est donc totalement adaptée à la réception d'un petit groupe d'auditeurs, venus écouter les cours de Marie Louise Dupiéry. L'espace domestique de cette savante s'ouvre donc au public, trois fois par semaine durant six semaines[56], en 1789 et 1790. À ces séances « intérieures » du midi viennent s'ajouter six séances en extérieur, désignées dans l'annonce du cours comme « quelques belles soirées qu'on choisira pour apprendre à connaître les étoiles »[57], et où elle « conduira ses auditeurs sur une terrasse pour leur montrer les constellations et les planètes »[58]. Quels étaient les espaces investis par Marie Louise Dupiéry et ses élèves pour réaliser ses observations ? À ce jour, aucune archive ne permet de répondre : très probablement un jardin parisien muni d'une terrasse, proche de l'appartement de la savante. La question de la sortie de nuit pour ce public de femmes n'est pas sans soulever la question de la bienséance. Ce frein aux déplacements nocturnes des femmes est ici en partie levé par le crédit accordé à la savante, veuve à cette époque et donc jouissant davantage de liberté que les jeunes femmes célibataires, et par la société non mixte que ces femmes constituent[59]. On peut mesurer l'importance de cette limite imposée aux femmes élèves, quelques années plus tard, grâce au témoignage de Victorine de Chastenay (1771-1855) qui assiste aux démonstrations publiques de François Arago (1786-1853) — son maître d'astronomie — à l'Observatoire où elle échange avec lui dans les années 1810[60]. Elle explique dans ses mémoires les difficultés à surmonter pour une femme (telles que la nécessité d'un chaperon, l'éloignement, la sécurité) :

> Malheureusement, on ne peut aller que le soir ou pendant la nuit rendre visite à ces astres, que j'adore toujours. Je ne pouvais aller seule au temple d'Uranie, je ne pouvais même y aller à pied avec un guide : le quartier est trop isolé. Maman m'interdisait même de m'y rendre en fiacre ; elle crut plusieurs fois me faire plaisir en m'y menant dans sa voiture, mais cette complaisance très grande de sa part, ne pouvait pas m'être agréable : je ne pouvais être certaine ni de l'heure, ni du jour. De plus, les entretiens que je venais chercher ne pouvaient pas, en présence de maman, avoir le caractère et le genre de portée que, sans me dire bien savante, ils auraient sans doute eu pour moi, si j'avais parlé seule. » (Chastenay, 1897, t. 2, p. 181)

---

[53] Concernant cette cérémonie, ainsi que les limitations temporelles et géographiques de l'enseignement public de Laura Bassi, se référer à (Findlen, 1993).
[54] « Cours d'astronomie ouvert principalement pour les dames, par Mme Du Piery », *Journal de Paris,* Supplément au numéro 120, 30 avril 1789, p. 549.
[55] AN/MC/ET/XXIII/771, Étude Richard, Inventaire après décès de Colin Alexandre Du Piéry, établi le 19 mai 1780.
[56] « Cours d'astronomie ouvert principalement pour les dames, par Mme Du Piery », *Journal de Paris,* Supplément au numéro 120, 30 avril 1789, p. 549-550.
[57] *Ibid.*

En tant que jeune femme célibataire privilégiée, échanger des propos savants en tête à tête avec un astronome, la nuit, à l'Observatoire de Paris est un acte condamné par les règles sociales genrées qui pourrait nuire gravement à sa réputation, et à la considération dont elle jouit. Ces règles, qui régissent les espaces publics comme privés constituent un frein majeur à l'investissement des femmes dans les sciences telles que l'astronomie.

---

[58] « Astronomie », *Gazette nationale ou le Moniteur universel*, n° 71, vendredi 12 mars 1790.

[59] Il est aussi très probable que ses « auditeurs » aient compté quelques hommes dans leurs rangs, mais en l'absence d'archives il n'est pas possible de se prononcer sur la constitution effective de son public, qui est considéré comme féminin dans l'annonce du cours.

[60] Cette pratique des démonstrations publiques est décrite dans (Aubin, Bigg & Sibum, 2010 ; Aubin, 2003 ; Wolf, 1902).

Ces exemples illustrant les pratiques privées comme publiques d'enseignement et d'apprentissage astronomique, voire plus largement de sciences, mettent en évidence le poids des règles genrées sur l'organisation de l'espace pratiqué par des femmes et sur la circulation des femmes oratrices, comme auditrices. Lorsqu'elles veulent enseigner ou apprendre les sciences, comme l'astronomie, elles sont contraintes la plupart du temps de se rendre dans des lieux dont l'accès est soumis à certaines règles bien plus nombreuses que pour les hommes. En tant que célibataires, elles ne peuvent se déplacer seules, et un chaperon leur est nécessaire afin de préserver leur réputation, rendant plus complexe toute démarche d'apprentissage par exemple. Les observations astronomiques, ayant lieu principalement de nuit, la bienséance les exclut systématiquement d'une pratique publique, sauf si elle s'effectue entre femmes. De plus, lorsqu'elles font publiquement acte d'un enseignement scientifique, leur corps est mis en scène à l'instar de celui de Marie Anne Pigeon d'Osangis ou de Laura Bassi, de manière à assurer leur intégration bienveillante par le monde savant.

**Conclusion**

Au XVIII$^e$ siècle, les savants et un très petit nombre de savantes circulent dans les mêmes espaces géographiques du savoir, qu'ils soient privés ou publics, domestiques ou institutionnels. Bien entendu, ces femmes rencontrent globalement beaucoup plus d'obstacles à investir les institutions et seules quelques-unes réussissent à le faire en France, de manière invisible ou officieuse, le seuil institutionnel restant infranchissable à cette période. Les contraintes liées à leur réputation et à leur sécurité limitent grandement leurs déplacements et leur liberté d'action dans le domaine savant, corsetant ainsi l'expression de leur goût des sciences. L'espace domestique, lieu de retrait, propice au travail intime, leur est indispensable, tout comme il l'est pour les hommes à cette époque. Le foyer participe de la production et de la circulation des savoirs scientifiques, et devient espace d'échange, de collaboration entre savants et savantes. L'organisation du temps et de quelques espaces mixtes s'y effectue de manière à optimiser la production savante, tout en respectant les obligations domestiques et mondaines. Cet espace, où peuvent se succéder pratiques privées et pratiques publiques de science, est régi par de nombreuses règles sociales, dont certaines sont genrées, imposant le lieu du domicile qui assure la bienséance pour une femme, les heures appropriées à ces pratiques, les personnes qui peuvent les partager (en lien avec leur crédit), le type de pratiques acceptables, etc. Dans l'espace domestique, ces femmes peuvent avoir les mêmes pratiques scientifiques que les hommes, voire plus rarement (en tout cas de manière visible) assumer des fonctions de responsabilité plus importantes comme dans l'atelier de Lalande. Elles administrent alors les espaces savants qui sont à leur portée en tant qu'intendantes scientifiques du savant. Parfois, l'exceptionnalité attribuée aux femmes par leurs pairs masculins, les révèle plus facilement à l'historien·ne ; et par là même les pratiques des « petites-mains », des subalternes (hommes comme femmes) — souvent peu documentées — deviennent accessibles. Les savoirs produits au foyer participent d'une dynamique plus globale de production et circulation des savoirs, parfois initiée par un savant ou une savante, parfois par une institution. Les va-et-vient entre espaces contribuent à la construction de nouveaux savoirs scientifiques, dans laquelle quelques femmes sont parties prenantes comme dans le cas assez exceptionnel des calculatrices de tables astronomiques.

L'existence d'une pratique sexuée des espaces, qu'ils soient domestiques ou institutionnels, apparaît clairement dans les cas étudiés ici. Plus les femmes y sont visibles (au travers d'interventions publiques par exemple), plus elles doivent se plier strictement aux règles sociales des Lumières afin de préserver leur réputation et leur crédit. Pour cela, leur corps « exposé » peut même constituer l'enjeu d'une véritable mise en scène. Celle-ci est d'ailleurs souvent associée à un discours genré (de la part d'admirateurs) présentant la savante ou l'intendante comme une femme dont la dévotion « quasi religieuse » aux sciences ne l'empêche pas de remplir ses devoirs familiaux. Lalande nous rappelle d'ailleurs concernant Nicole Reine Lepaute que :

> ses calculs ne l'empêchaient point de s'occuper des affaires de la maison, les livres de commerce étaient à côté des tables astronomiques ; le goût et l'élégance étaient dans ses ajustements, sans nuire à ses études. Son mari avait pour elle cette considération qui tient du respect [...] Elle était cependant remplie de prévenances pour lui ; elle le servait avec empressement. » (Lalande, 1803, p. 680)

Une femme savante, oui, mais avant tout élégante et raffinée, dévouée à son époux et à son service !

Isabelle LEMONON-WAXIN

## Références

ANCELIN Justine (2011), *Science, académisme et sociabilité savante. Édition critique et étude du Journal de la vie privée de Jean-Dominique Cassini (1710-1712),* Thèse de l'École nationale des chartes, Paris.

AUBIN David, BIGG Charlotte & SIBUM H. Otto (2010), *The Heavens on Earth: Observatories and Astronomy in Nineteenth-Century Science and Culture,* Durham, Duke University Press.

AUBIN David (2003), « The Fading Star of the Paris Observatory in the Nineteenth Century: Astronomers' Urban Culture of Circulation and Observation », *Osiris*, vol. 18, p. 79-100, doi : 10.1086/649378

BADINTER Élisabeth (2000), *Émilie, Émilie : l'ambition féminine au XVIIIe siècle*, Paris, Flammarion.

BADINTER Élisabeth (2004-2005), « Un couple d'astronomes : Jérôme Lalande et Reine Lepaute », *Société archéologique, scientifique et littéraire de Béziers,* sér. 10, vol. 1, p. 71-76.

BOISTEL Guy (2001), *L'astronomie nautique au XVIIIe siècle en France : tables de la lune et longitudes en mer,* Lille, ANRT.

BOISTEL Guy (2004), « Nicole Lepaute et l'hortensia », *Cahiers Clairaut*, vol. 108, p. 13-17.

BOISTEL Guy (2007), « Jérôme Lalande, premier astronome médiatique », *Les Génies de la science,* vol. 32, p. 10-13.

BOISTEL Guy, LAMY Jérôme & LE LAY Colette (2010), *Jérôme Lalande, 1732-1807 : une trajectoire scientifique*, Rennes, Presses universitaires de Rennes.

BROSSARD J. (1895), « Quelques lettres inédites de Lalande », *Annales de la Société d'émulation d'agriculture, lettres et arts de l'Ain*, p. 66-86.

CERTEAU Michel de (1980), *L'invention du quotidien, Tome 1. Arts de faire*, Paris, Union générale d'éditions.

CHASTENAY Victorine de (1897), *Mémoires de madame de Chastenay, 1771-1815 : L'empire. La restauration. Les cent-jours,* Tome 2, Alphonse Roserot (éd.), Paris, Plon.

CROWSTON Clare Haru (2013), *Credit, Fashion, Sex: Economies of Regard in Old Regime France*, Durham, Duke University Press.

DEIAS Dalia (2020), *Inventer l'Observatoire : sciences et politiques sous Giovanni Domenico Cassini (1625-1712)*, Thèse de doctorat, École des hautes études en sciences sociales, Paris.

DUBY Georges & Michelle PERROT (1991), *Histoire des femmes en Occident*, Paris, Plon.

DUMONT Simone (2007), *Un astronome des Lumières : Jérôme Lalande*, Paris, Observatoire de Paris.

DUPIERY Marie Louise (1802), *Table alphabétique et analytique des matières contenues dans les dix tomes du Système des connaissances chimiques rédigée par M^me Dupiéry et revue par le C^en Fourcroy*, Paris, Baudouin.

ELIAS Norbert (2016), « L'espace privé. Privatraum ou privater Raum ? », *Socio*, vol. 7, p. 25-37, doi : 10.4000/socio.2369

FAGES Volny (2012), *Les origines du monde : cosmogonies scientifiques en France (1860-1920) : acteurs, pratiques, représentations*, Thèse de doctorat, École des hautes études en sciences sociales, Paris.

FERLAND Catherine & GRENIER Benoit (2013), « Quelque longue que soit l'absence : procurations et pouvoir féminin à Québec au XVIII^e siècle », *Clio. Femmes, Genre, Histoire*, vol. 37, p. 197-225, doi : 10.4000/clio.11053

FINDLEN Paula (1993), « Science as a Career in Enlightenment Italy: the Strategies of Laura Bassi », *Isis*, vol. 84, n° 3, p. 441-469, doi : 10.1086/356547

FINNEGAN Diarmid A. (2008), « The Spatial Turn: Geographical Approaches in the History of Science », *Journal of the History of Biology*, vol. 41, n° 2, p. 369-388, doi : 10.1007/s10739-007-9136-6

FONTAINE Laurence (2008), *L'économie morale : pauvreté, crédit et confiance dans l'Europe préindustrielle*, Paris, Éditions Gallimard.

FOURCROY Antoine-François (1810), *Catalogue des livres de la bibliothèque de feu M. A. F. de Fourcroy*, Paris, Tilliard.

HABERMAS Jurgen (1962), *L'espace public : archéologie de la publicité comme dimension constitutive de la société bourgeoise*, Paris, Payot.

HAFTER Daryl M. & KUSHNER Nina (2015), *Women and Work in Eighteenth-Century France*, Bâton Rouge, LSU Press.

HOTTIN Christian (2011), « Retour sur un patrimoine parisien méconnu : les espaces de transmission du savoir à l'époque moderne (II). Naissance d'une architecture : quatre projets exceptionnels (ca. 1760 – ca. 1790) », *In Situ*, vol. 17, En ligne : https://journals.openedition.org/insitu/1069

JACOB Christian (2014), *Qu'est-ce qu'un lieu de savoir ?*, Marseille, OpenEdition Press.

KWAN Alistair Marcus (2010), *Architectures of Astronomical Observation: From Sternwarte Kassel (circa 1560) to the Radcliffe Observatory (1772)*, New Haven, Yale University.

Isabelle LEMONON-WAXIN

GALISON Peter (1999), « Buildings and the Subject of Science », dans Peter GALISON & Emily Ann THOMPSON (éds.), *The Architecture of Science,* Cambridge, MIT Press, p. 1-28.
LALANDE Jérôme (1801), *Histoire céleste française*, Imprimerie de la République, Paris.
LALANDE Jérôme (1803), *Bibliographie astronomique : avec l'histoire de l'astronomie depuis 1781 jusqu'à 1802,* Imprimerie de la République, Paris.
LALANDE Jérôme (2007), *Lalandiana I : Lettres à Mme Du Pierry et au juge Honoré Flaugergues,* Jean-Claude PECKER & Simone DUMONT (éds.), Paris, Vrin.
LALANDE Jérôme (2014), *Lalandiana II : Mission à Berlin. Lettres à Jean III Bernoulli et à Elert Bode,* Jean-Claude PECKER & Simone DUMONT (éds.), Paris, Vrin.
LAMY Jérôme (2007), *L'observatoire de Toulouse aux XVIII$^e$ et XIX$^e$ siècles : archéologie d'un espace savant*, Rennes, Presses universitaires de Rennes.
LAUNAY Françoise (2015), « Le fabuleux destin de Marie Jeanne Harlay », *L'Astronomie*, vol. 129, n° 88, p. 37-43.
LÉMONON-WAXIN Isabelle (2016), « Gender and Space in Enlightenment Science: Madame Dupiéry's Scientific Work and Network », dans Donald L. OPITZ, Staffan BERGWIK & Brigitte van TIGGELEN (éds.), Basingstoke, Palgrave Macmillan, p. 41-60.
LEMONON-WAXIN Isabelle (2019), *La Savante des Lumières françaises, histoire d'une persona : pratiques, représentations, espaces et réseaux*, Thèse de doctorat, École des hautes études en sciences sociales, Paris.
LEPAUTE Jean André (1755), *Traité d'horlogerie contenant tout ce qui est nécessaire pour bien connoitre et pour régler les pendules et les montres*, Paris, Chardon.
LILTI Antoine (2015), « Le pouvoir du crédit au XVIII$^e$ siècle. Histoire intellectuelle et sciences sociales », *Annales. Histoire, Sciences Sociales*, vol. 70, n° 4, p. 957-978, doi : 10.3917/anna.704.0957
LIVINGSTONE David N. (2003), *Putting Science in Its Place: Geographies of Scientific Knowledge*, Chicago / Londres, University of Chicago Press.
MOMMERTZ Monika (2005), « The Invisible Economy of Science. A New Approach to the History of Gender and Astronomy at the Eighteenth-Century Berlin Academy of Science », dans Judith P. ZINSSER (éd.), *Men, Women, and the Birthing of Modern Science*, DeKalb, Northern Illinois University Press, p. 159-178.
OPHIR Adi & SHAPIN Steven (1991), « The Place of Knowledge. A Methodological Survey », *Science in context*, vol. 4, n° 1, p. 3-22.
OPITZ Donald L., BERGWIK Staffan & VAN TIGGELEN Brigitte (éds.) (2015), *Domesticity in the Making of Modern Science,* Basingstoke, Palgrave Macmillan.

ROCHE Daniel (1969), « Un savant et sa bibliothèque au XVIIIe siècle. Les livres de Jean-Jacques Dortous de Mairan, secrétaire perpétuel de l'Académie des sciences, membre de l'Académie de Béziers », *Dix-huitième siècle*, vol. 1, p. 47-88, doi : 10.3406/dhs.1969.879

ROGERS Rebecca (2000), « Le sexe de l'espace : réflexions sur l'histoire des femmes aux XVIIIe-XXe siècles dans quelques travaux américains, anglais et français », dans Jean-Claude WAQUET, Odile GOERG & Rebecca ROGERS (éds.), *Les espaces de l'historien*, Strasbourg, Presses universitaires de Strasbourg, p. 181-202.

SCHIEBINGER Londa (1987), « Maria Winkelmann at the Berlin Academy: A Turning Point for Women in Science », *Isis*, vol. 78, n° 2, p. 174-200.

SHAPIN Steven (1988), « The House of Experiment in Seventeenth-Century England », *Isis*, vol. 79, n° 3, p. 373-404.

SHAPIN Steven (1998), « Placing the View from Nowhere: Historical and Sociological Problems in the Location of Science », *Transactions of the Institute of British Geographers*, vol. 23, n° 1, p. 5-12, doi : 10.1111/j.0020-2754.1998.00005.x

TERRALL Mary (2014), *Catching Nature in the Act: Réaumur and the Practice of Natural History in the Eighteenth Century*, Chicago, University of Chicago Press.

TILLY Louise A. & SCOTT Joan W. (2016), *Women, Work and Family*, Londres, Routledge.

WITHERS Charles W. J. (2009), « Place and the "Spatial Turn" in Geography and in History », *Journal of the History of the Ideas*, vol. 70, n° 4, p. 637-658.

WOLF Charles (1902), *Histoire de l'Observatoire de Paris de sa fondation à 1793*, Paris, Gauthier-Villars.